\documentclass{article}
\usepackage{subcaption}
\usepackage{graphicx}
\usepackage{amsmath}
\usepackage{amssymb}
\usepackage{url}
\usepackage[round]{natbib}
\title{A multi-scale loss formulation for learning a probabilistic model with proper score optimisation}

\author{Simon Lang, Martin Leutbecher, and Pedro Maciel\\
\\
ECMWF, Reading, UK}
\date{12 June 2025}

\begin{document}

\maketitle

\begin{abstract}
We assess the impact of a multi-scale loss formulation for training probabilistic machine-learned weather forecasting models. The multi-scale loss is tested in AIFS-CRPS, a machine-learned weather forecasting model developed  at the European Centre for Medium-Range Weather Forecasts (ECMWF). AIFS-CRPS is trained by directly optimising the almost fair continuous ranked probability score (afCRPS). The multi-scale loss better constrains small scale variability without negatively impacting forecast skill. This opens up promising directions for future work in scale-aware model training.
\end{abstract}

\section{Introduction}
Over the last few years, probabilistic machine-learned weather prediction models have begun to rival physics-based numerical weather prediction (NWP) systems in skill \citep{kochkov2024neural, price2023gencast, ensemble_blog, lang2024aifs-crps}. AIFS-CRPS \citep{lang2024aifs-crps} is based on the machined-learned weather forecasting model AIFS \citep{lang2024aifs}, developed at the European Centre for Medium-Range Weather Forecasts (ECMWF). AIFS-CRPS produces skilful predictions by directly optimising a score based on a proper scoring rule, the almost fair continuous ranked probability score (afCRPS). The model learns to shape Gaussian noise to represent uncertainty in the atmospheric state and achieves ensemble forecast skill that is competitive with, or superior to, the physics-based IFS ensemble \citep{molteni1996ecmwf, leutbecher2008ensemble, lang2021more, lang81380} at ECMWF.

The afCRPS loss function used in AIFS-CRPS is computed point-wise on the full output field. 
However, atmospheric processes are inherently multi-scale, and different scales contribute to a different degree to the loss function. The  scale-dependent verification of ensemble forecasts has been explored with wavelets and spectral band pass filters by \cite{casati.wilson.2007,jung.leutbecher.2008}, respectively. These studies have shed additional light on the skill of ensemble forecasts as function of spatial scale. In contrast, the standard application of scoring rules to gridded forecasts does not take the spatial scale of the forecast errors and ensemble perturbations into account. An exception is the work of \cite{kochkov2024neural} who incorporate a spectral CRPS terms in their loss function to train a hybrid model that combines a differentiable solver for atmospheric dynamics with a machine-learned physics module.

The question that arises is whether the machine-learned forecast model AIFS-CRPS can be improved by adding additional constraints via a loss function that evaluates different spatial scales separately. Here, we test the effect of adding a multi-scale component to the afCRPS training objective. We evaluate whether this modification leads to improved spatial structures compared to the scale-unaware loss function.

\section{Methodology}
\subsection{The multi-scale loss}
We consider predictions and targets which are scalar functions on an $\ell$-dim manifold $\mathcal{M}$
\begin{displaymath}
   \phi: \mathcal{M}  \rightarrow \mathbb{R}.
\end{displaymath}
Later in this section, an idealized 1-dim example will be shown. The remainder of the paper focusses on an application to the 2-sphere for global weather prediction.  However, the concepts are generic and can be applied to higher dimensions $\ell>2$ as well.
In most applications, we expect that discretization of these functions on suitable grids will be used. For what follows, we will not distinguish explicitly between the continuous case and the discretized case and will adopt a lightweight notation. Spatial integration over the manifold will be denoted by integrals with the understanding that these are replaced by suitable finite sums over grids for the discrete case. 

Consider an optimisation of probabilistic predictions for a target using a scoring rule $\mathcal{S}$ for scalars.  
Let $x_j: \mathcal{M}  \rightarrow \mathbb{R}$ and $y: \mathcal{M}  \rightarrow \mathbb{R}$ denote the $j$-th prediction and the target, respectively. Then a loss can be defined as 
\begin{equation} 
\label{eq:scale-unaware-loss}
    \mathcal{L} = c \int_{\mathcal{M}} \mathcal{S} ( [x_{j} \,|\; j=1,\ldots M], y) \,\mathrm{d}\mu
\end{equation}
The score $\mathcal{S}$ is computed for each location $q\in\mathcal{M}$ and then spatially averaged. Here $\mu$ denotes a measure on $\mathcal{M}$ and $c$ is a normalisation constant. The loss is not scale-aware as the scoring rule depends only  on the marginal distributions sampled by the ensemble of predictions at each location $q\in\mathcal{M}$. In order to distinguish this loss from the multi-scale loss introduced next, we will refer to it by $\mathcal{L}_{\text{scale-unaware}}$.

Now, a multi-scale loss will be introduced based on a sequence of ordered smoothing operators \( D_i,\quad i=1,\ldots, n-1 \) which remove the smaller scales of functions $\phi: \mathcal{M} \rightarrow \mathbb{R}$. It is assumed that $D_{i}$ smooths more than $D_{i+1}$. These operators induce a partition of a function $\phi$ on the manifold into $n$ scales
\begin{align}
    \phi_{\text{scale}\,1} &= D_1( \phi) \nonumber \\
    \phi_{\text{scale}\,2} &= D_2( \phi) -  D_1( \phi)\nonumber \\
    &\;\vdots  \nonumber\\
    \phi_{\text{scale}\,n} &= \phi  - D_{n-1}( \phi) \nonumber 
\end{align}
Then, the $n$-scale loss is defined as a weighted sum of the loss for each scale $i$
\begin{equation}
\mathcal{L}_{n\text{-scale}} = \sum_{i=1}^n \zeta_i \,c\int_{\mathcal{M}}\mathcal{S}( [x_{j,\mathrm{scale}\,i } \,|\; j=1,\ldots M], y_{\mathrm{scale}\,i}) \,\mathrm{d}\mu    \label{eq:multi-scale-loss}
\end{equation}
with  weight $\zeta_i>0$ for scale $i$. It is straightforward to introduce as many loss scales as required. The $D_i$ could be implemented as linear kernel smoothers with width decreasing with $i$. Alternatively, a spectral filter could be used if spectral transforms are available for the manifold $\mathcal{M}$. 
 
\subsubsection{Simulation study with monochromatic waves}
\begin{figure}
    \centering
    \includegraphics[width=0.9\linewidth]{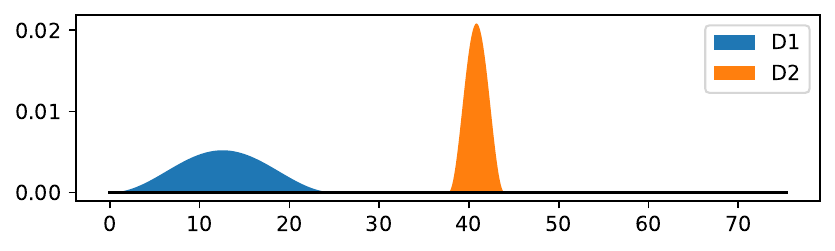}
    \caption{Smoothing kernels for the 1-dim simulation study}
    \label{fig:kernels1d}
\end{figure}

\begin{figure}
    \centering
    \includegraphics[width=.9\linewidth]{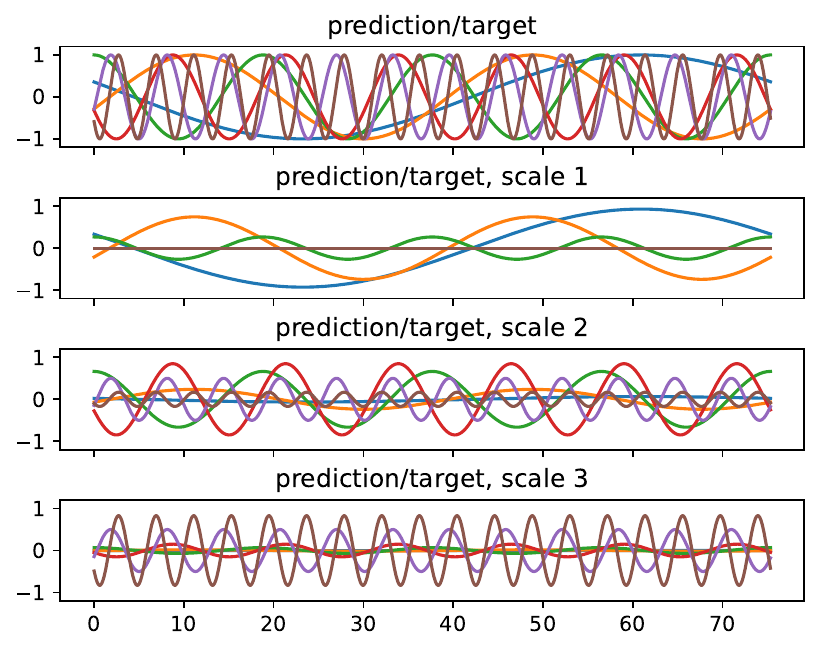}
    \caption{Predictions and targets are sine waves with random phases. The three scales focus on different wavelengths: Scale~1 is dominated by the largest wavelengths, scale~2 emphasises intermediate wavelengths and scale~3  the shortest wavelengths.}
    \label{fig:waves1d}
\end{figure}

In order to motivate why a multi-scale loss may be useful, this section illustrates the concept with a one-dimensional simulation study on a periodic domain. The predictions and the target are sine waves with a random phase and unit amplitude. We consider a 3-scale loss defined with the kernels shown in Figure~\ref{fig:kernels1d}
The implied partition of the waves into different spatial scales is shown for a couple of different waves in Figure~\ref{fig:waves1d}

For the simulation study, we use the fair CRPS as scoring rule. It accounts for the finite number of predictions and estimates the CRPS one would obtain with probabilities estimated from an infinite sample. We consider as target waves with wavenumber $k_t$. 
Then, for a set of wavenumbers $k$ we consider predictions with an ensemble size of $M=8$. The scale-unaware loss and the 3-scale loss are computed for each wavenumber. This is repeated for 4000 realisations of the truth and the predictions. 

\begin{figure}
    \centering
    \includegraphics[height=0.25\linewidth]{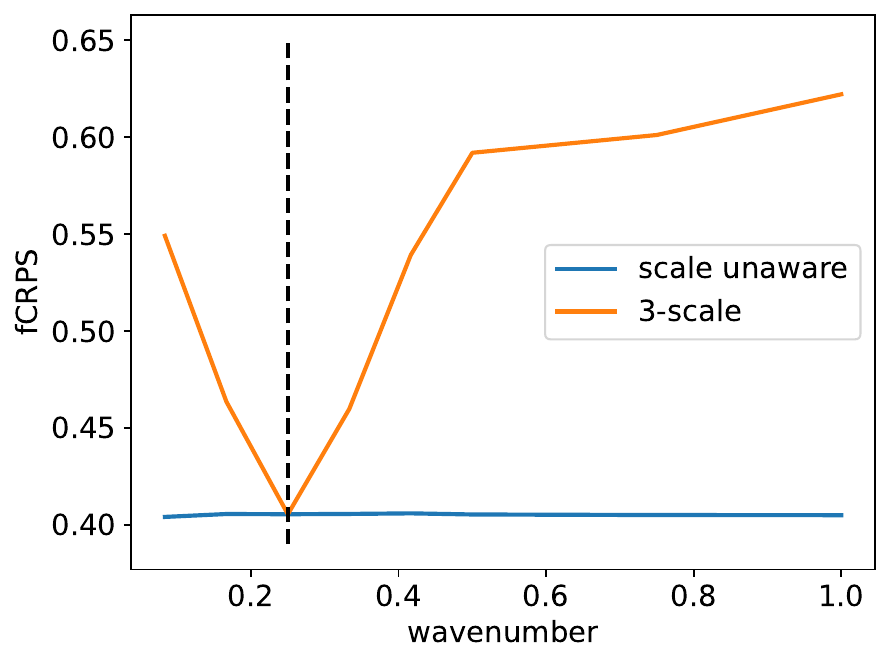}%
    \includegraphics[height=0.25\linewidth]{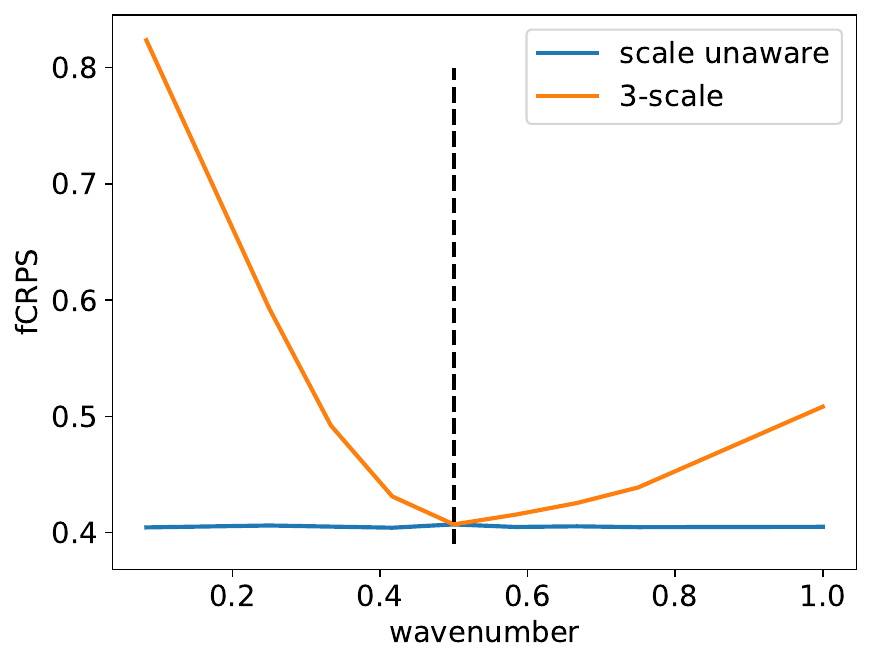}%
    \includegraphics[height=0.25\linewidth]{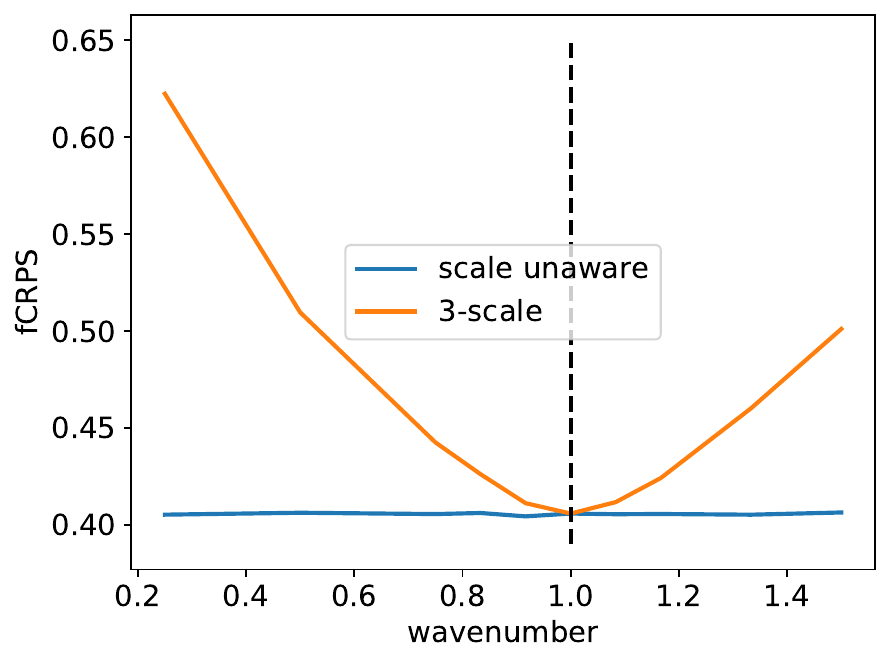}
    \caption{Fair CRPS as function of the predicted wavenumber for the scale-unaware loss and the three-scale loss. The dashed vertical line indicates the wavenumber of the target. The three panels consider target wavenumbers of $\frac{1}{4}, \frac{1}{2}$  and 1.}
    \label{fig:loss1d}
\end{figure}

Figure~\ref{fig:loss1d} shows that the scale-unaware loss is indeed invariant of the wavenumber of the prediction. In contrast, the 3-scale loss  correctly identifies the target wavenumber.

\subsection{AIFS-CRPS experiments}
We compare two versions of AIFS-CRPS, one reference experiment, trained with the scale-unaware loss formulation and one with the multi-scale loss formulation. The models have a spatial resolution of approximately $0.25^\circ$ \citep[N320 reduced Gaussian grid, ][]{Wedi2014}.

\subsubsection{Loss objective}
We use the almost fair Continuous Ranked Probability Score (afCRPS, \citet{lang2024aifs-crps}) with $\alpha=0.95$ as the loss objective. The afCRPS is a linear combination of the CRPS \citep{hersbach2000} and the fair version of the CRPS (fCRPS, \citet{Ferro2013,leutbecher2019}). 
The fair score enables training with small ensembles sizes as low as two members. However, it exhibits a degeneracy which leaves one member unconstrained if all other members are identical to the observed value. The degeneracy can be addressed by using the almost fair CRPS introduced by
 \cite{lang2024aifs-crps}:
\begin{align}
\text{afCRPS}_{\alpha} &:= \alpha \, \text{fCRPS} + (1-\alpha) \text{CRPS} \nonumber \\
&= \frac{1}{M} \sum_{j=1}^{M} |x_j - y| - \frac{M - 1 + \alpha}{2M^2(M-1)} \sum_{j=1}^{M} \sum_{k=1}^{M} |x_j - x_k| \nonumber \\
&= \frac{1}{M} \sum_{j=1}^{M} |x_j - y| - \frac{1 - \epsilon}{2M(M-1)} \sum_{j=1}^{M} \sum_{k=1}^{M} |x_j - x_k| \nonumber
\label{afcrps_original}
\end{align}
with $\epsilon := \frac{(1-\alpha)}{M}$. Here, the $x_j$ and $y$ denote ensemble forecasts and the analysis, respectively. 

The scale-unaware loss is defined consistently with \cite{lang2024aifs-crps} using the almost fair CRPS as scoring rule $\mathcal{S}$ in \eqref{eq:scale-unaware-loss}.  Likewise, a multi-scale loss is defined by using the almost fair CRPS as scoring rule in \eqref{eq:multi-scale-loss}.  We focus on a two-scale version and use equal weighting for both scales,  $\zeta_i=1$. 

\begin{figure}[htpb]
\centering
\begin{subfigure}{0.9\linewidth}
    \caption{target}\label{fig:org}
    \includegraphics[width=\linewidth]{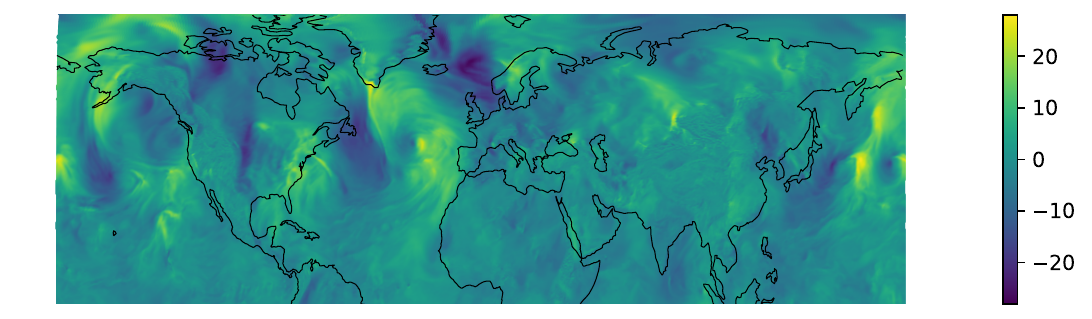}
\end{subfigure}
\begin{subfigure}{0.9\linewidth}
    \caption{target, scale~1}\label{fig:filtered}
    \includegraphics[width=\linewidth]{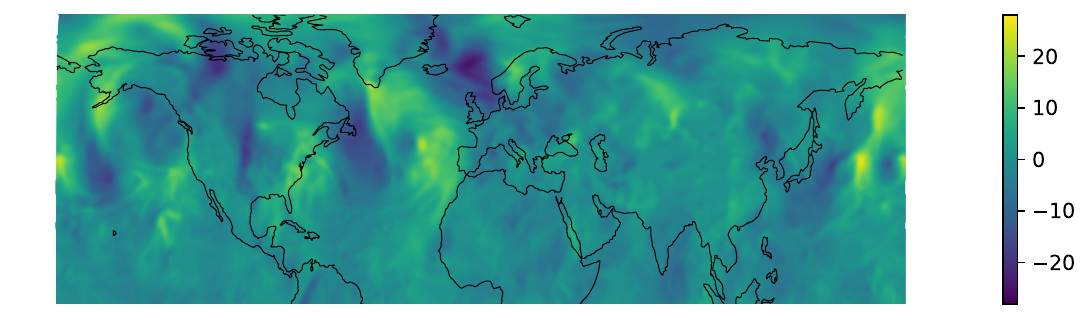}
\end{subfigure}
\begin{subfigure}{0.9\linewidth}
    \caption{target, scale~2}\label{fig:diff}
    \includegraphics[width=\linewidth]{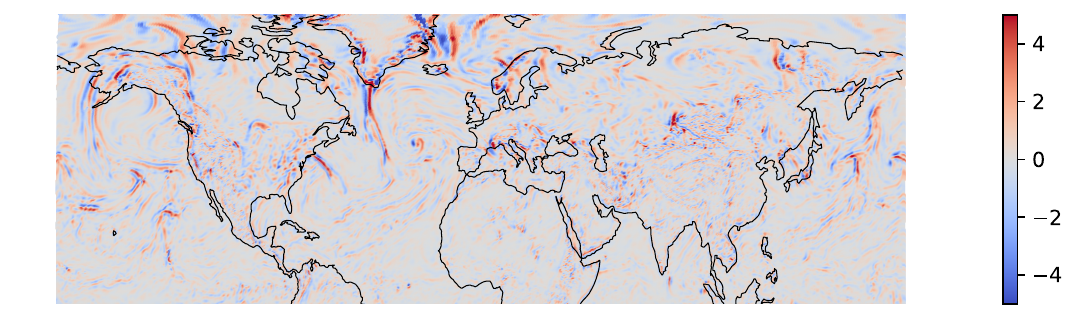}
\end{subfigure}
\caption{ERA5 v-component of wind (in $\mathrm{m\,s^{-1}}$) at 850~hPa, full field (\subref{fig:org}), field after filtering  (\subref{fig:filtered}) and difference between filtered and full field (\subref{fig:diff}).}
\label{fig:filtering}
\end{figure}

The smoothing operator is a linear filter implemented via sparse matrix multiplication. We use a Gaussian kernel, which is easily parametrised, though other filters could be employed. Its standard deviation is set to eight times the grid spacing. The kernel weights are linearly normalised to sum to one. We use matrices generated with ECMWF's Meteorological Interpolation and Regridding (MIR) software package \citep{maciel2017mir}. An example of a target field and its two scales is shown in Figure~\ref{fig:filtering}.

\subsubsection{Training}
The training and model set-up follows \citet{lang2024aifs-crps}. However, here we do not fine tune on operational analyses data of the physics-based Integrated Forecasting System (IFS). Instead we only train on the Copernicus ERA5 reanalysis dataset produced by ECMWF \citep{hersbach2020era5}. The training consists of three sequential stages. In the first stage, the model learns to forecast a single 6-hour time step ahead. In the second stage, we extend training to an auto-regressive setup with two 6-hour forecast steps. The third stage involves training with progressively longer forecast windows. The forecast length is increased after each epoch by 6~hours, up to a maximum of 72 hours. For the multi-scale loss version of the AIFS-CRPS, we start from the scale-unaware loss pre-trained model after the first training stage and then train the model with the multi-scale loss during stage 2 and 3.

The first stage comprises 300,000 parameter updates, starting with an initial learning rate of $10^{-3}$. We apply a cosine learning rate schedule with 1,000 warm-up steps, during which the learning rate increases linearly from zero to its maximum value, then gradually decreases back to zero. The second stage involves 60,000 iterations, using a cosine schedule with 100 warm-up steps and an initial learning rate of $ 10^{-5}$. The third stage includes approximately 45,000 iterations, with a fixed learning rate ($ 10^{-6}$). We use a batch size of 16 throughout training. The AdamW optimizer (\cite{loshchilov2018decoupled}) is used with $\beta$-coefficients of 0.9 and 0.95, and a weight decay  of 0.1. The training data consists of the  years 1979 to 2017 and the year 2018 is reserved for validation.

\section{Results}
To assess the impact of the multi-scale loss we compare 8 member ensemble forecasts of the scale-unaware loss and multi-scale loss trained AIFS-CRPS version. Forecasts have been run for each day of 2019, initialised at 00 UTC. Forecasts are started from ERA5 analyses and with initial perturbations derived from the ERA5 ensemble of data assimilations. 

\begin{figure}[htpb]
\centering
\begin{subfigure}{0.49\linewidth}
    \caption{500\,hPa geopotential}\label{fig:fcrpsz500}
    \includegraphics[width=\linewidth]{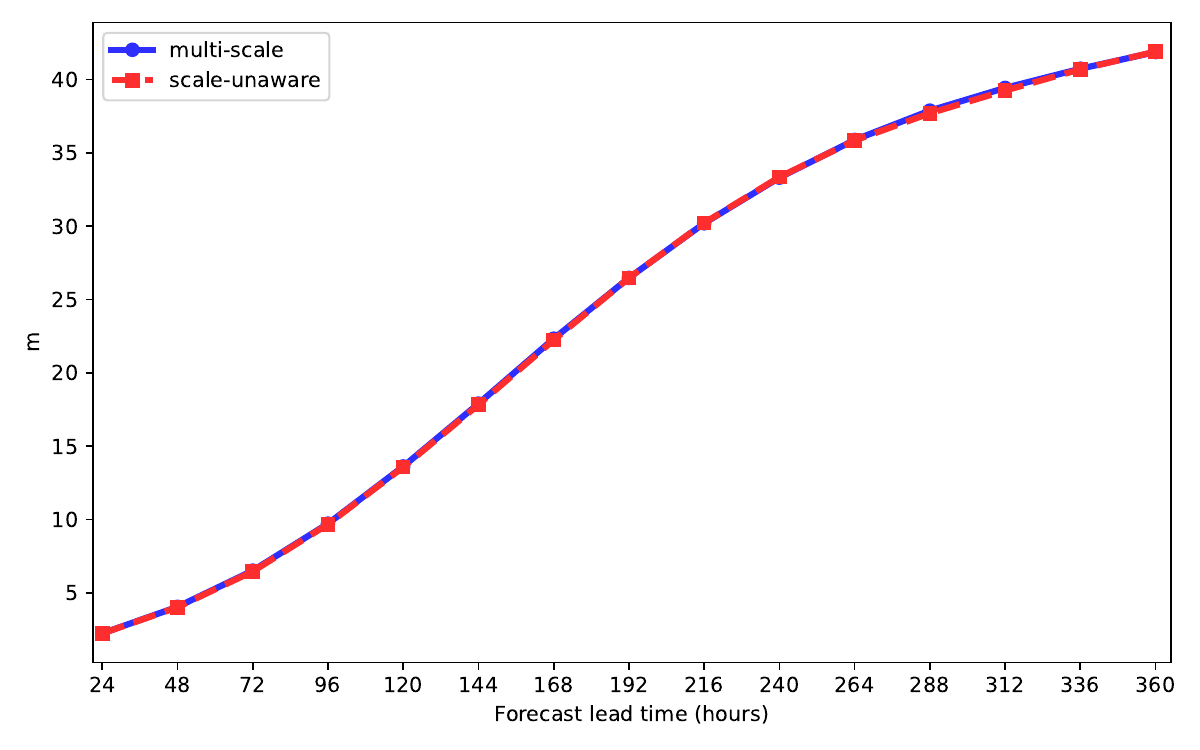}
\end{subfigure}
\hfill
\begin{subfigure}{0.49\linewidth}
    \caption{850\,hPa temperature}\label{fig:fcrpst850}
    \includegraphics[width=\linewidth]{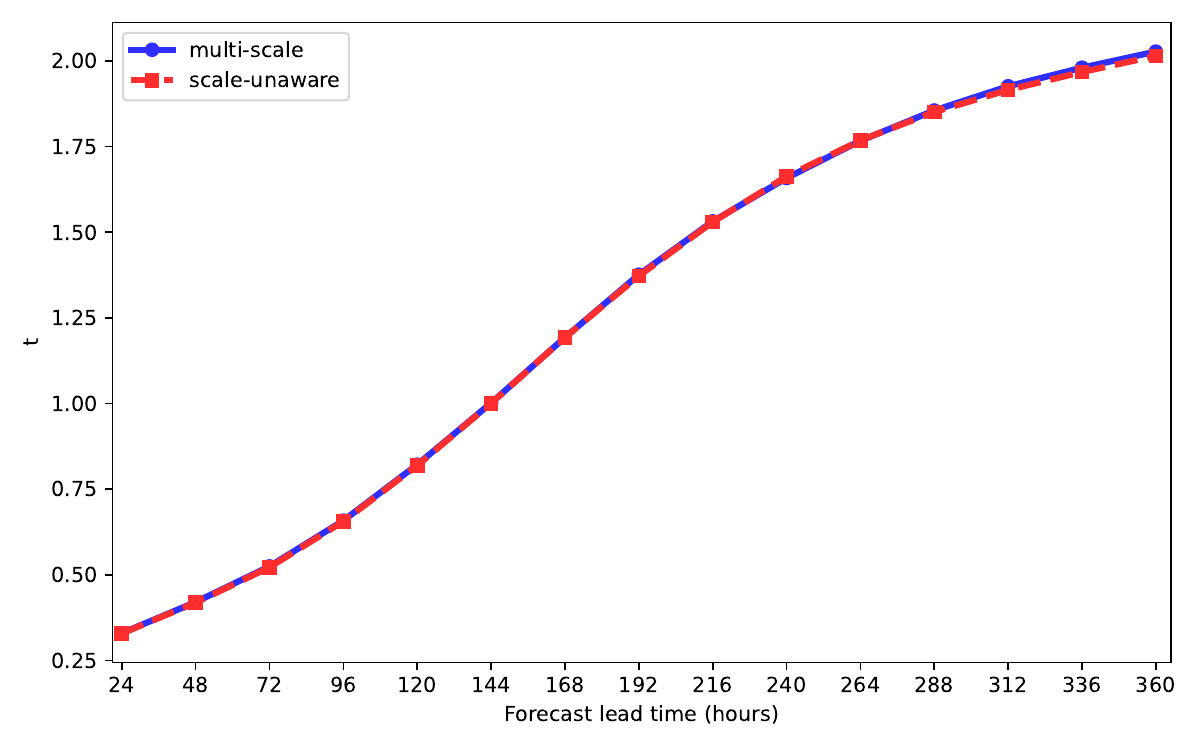}
\end{subfigure}
\begin{subfigure}{0.49\linewidth}
    \caption{850\,hPa windspeed}\label{fig:fcrpsff850}
    \includegraphics[width=\linewidth]{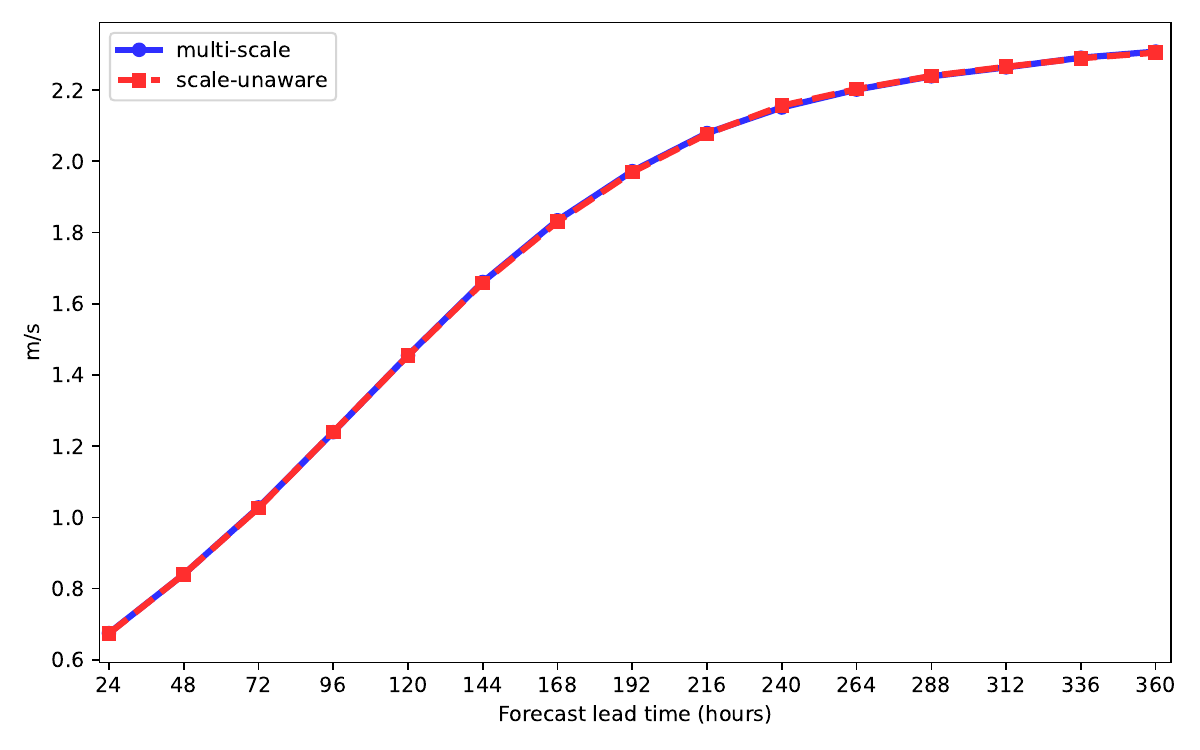}
\end{subfigure}
\caption{Fair CRPS for (\subref{fig:fcrpsz500}) northern hemisphere geopotential at 500~hPa, (\subref{fig:fcrpst850}) temperature at 850~hPa  and (\subref{fig:fcrpsff850}) windspeed at 850~hPa. Scores are shown for the year 2019, forecasts are initialised on 00 UTC each day. Fields are interpolated to a $1.5^\circ$ grid for verification, following standard practice.}
\label{fig:scores}
\end{figure}

Both experiments exhibit equal skill, as demonstrated by the nearly identical skill scores. Figure~\ref{fig:scores} shows examples for the Northern Hemisphere extra-tropics, where the curves are on top of each other. The same is true across other variables and regions (not shown).

\begin{figure}[htpb]
\centering
\begin{subfigure}{0.7\linewidth}
    \caption{}\label{fig:singleglob}
    \includegraphics[width=\linewidth, trim={2.18cm 3.5cm 11.19cm 8cm}, clip]{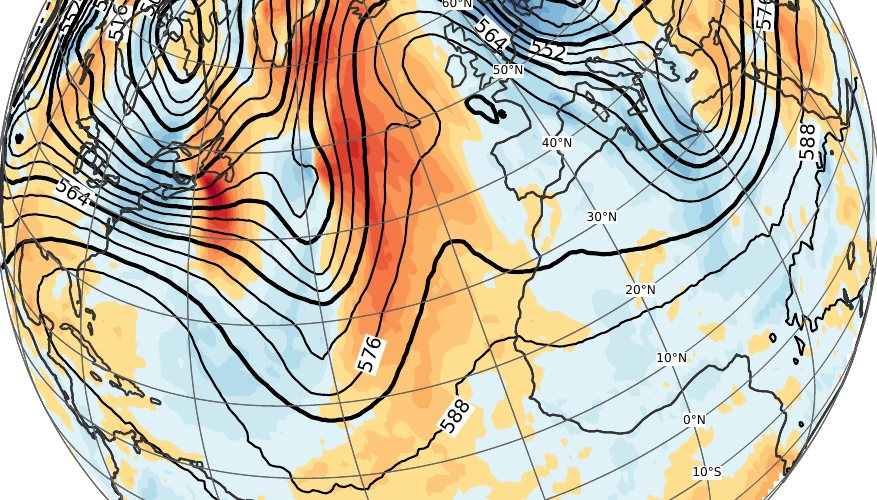}
\end{subfigure}
\begin{subfigure}{0.7\linewidth}
    \caption{}\label{fig:mscaleglob}
    \includegraphics[width=\linewidth, trim={2.18cm 3.5cm 11.19cm 8cm}, clip]{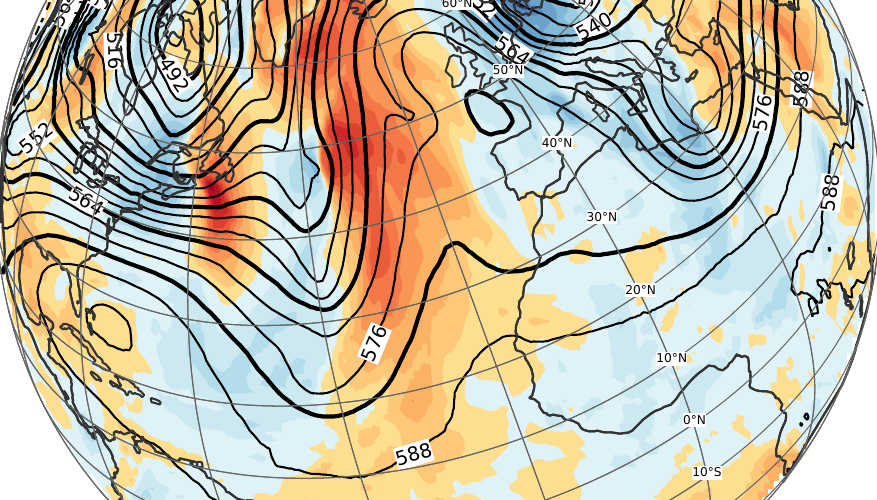}
\end{subfigure}
\begin{subfigure}{0.7\linewidth}
    \caption{}\label{fig:era5glob}
    \includegraphics[width=\linewidth, trim={2.18cm 3.5cm 11.19cm 8cm}, clip]{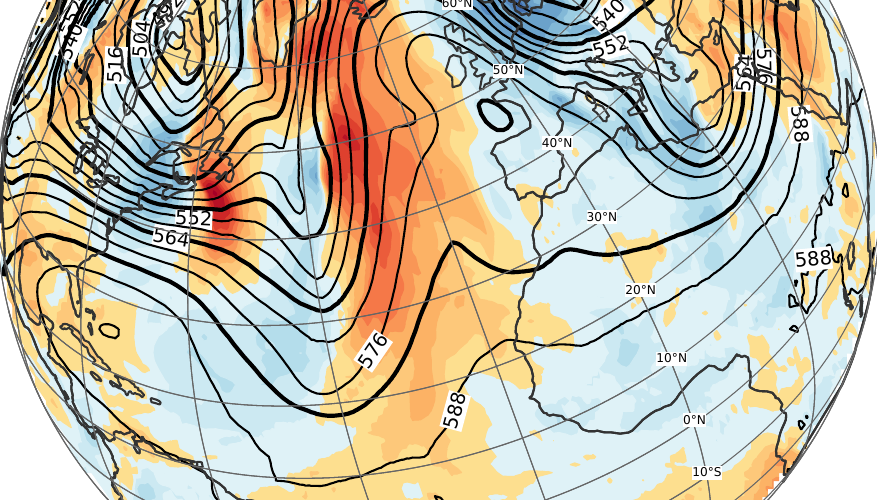}
\end{subfigure}
\caption{Geopotential at 500~hPa (dam, contours) and v-component of wind at 850~hPa (shaded) of (\subref{fig:singleglob}) the scale-unaware loss experiment, (\subref{fig:mscaleglob}) the multi-scale loss experiment  and (\subref{fig:era5glob}) the ERA5 analysis. Shown are 24~h forecasts of member 1, initialised on 2019-01-01 00 UTC (\subref{fig:singleglob}, \subref{fig:mscaleglob}) and verifying analysis on 2019-01-02 00 UTC (\subref{fig:era5glob}).}
\label{fig:case}
\end{figure}

However, examining individual forecast fields reveals that the scale-unaware loss experiment contains more small-scale variability than the multi-scale loss experiment and  the ERA5 analysis. This is evident in regions with weaker gradients, where contour lines appear more variable. For example, the 12-hour forecast 588 dam isohypse in figure~\ref{fig:case} shows noticeable differences between the scale-unaware loss experiment (figure~\ref{fig:singleglob}), and the multi-scale loss experiment (figure~\ref{fig:mscaleglob}) and the ERA5 analysis (figure~\ref{fig:era5glob}). On the other hand, The multi-scale experiment and the ERA5 analysis are in better agreement.

\begin{figure}[htpb]
\centering
\begin{subfigure}{0.8\linewidth}
    \caption{}\label{fig:spc_z_1}
    \includegraphics[width=\linewidth]{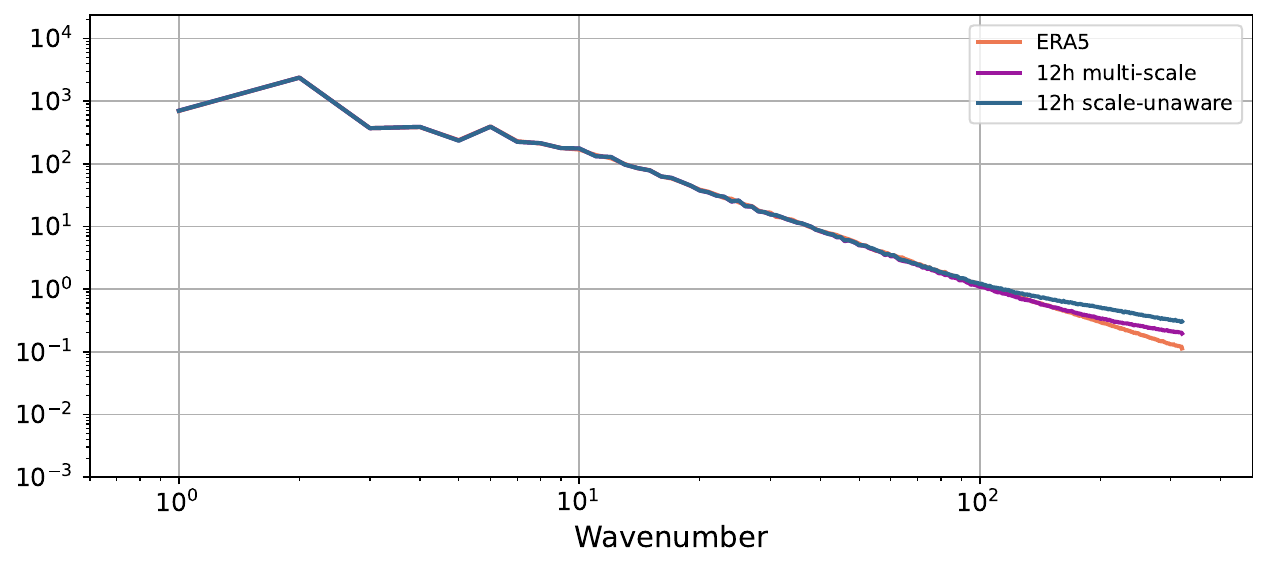}
\end{subfigure}
\begin{subfigure}{0.8\linewidth}
    \caption{}\label{fig:spc_t_1}
    \includegraphics[width=\linewidth]{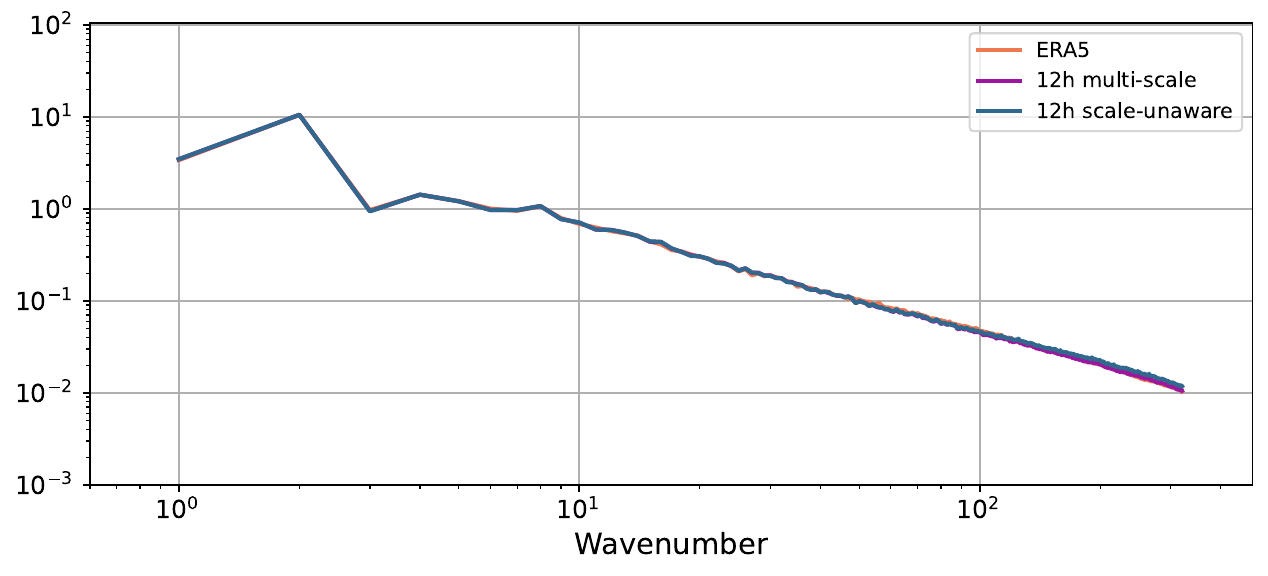}
\end{subfigure}
\caption{Spectra of (\subref{fig:spc_z_1}) geopotential at 500~hPa  and (\subref{fig:spc_t_1}) temperature at 850~hPa  for the experiment trained with the multi-scale loss and the experiment trained with the scale-unaware loss for 12~h forecasts. ERA5 refers to the initial conditions. The spectra are averaged over 11 initial dates for a single ensemble member.}
\label{fig:spec1}
\end{figure}

Spectra of forecast fields show that the multi-scale loss experiment better constrains small-scale variability compared to the scale-unaware loss experiment, when compared to the ERA5 analysis (figure~\ref{fig:spec1}). The impact is most pronounced for fields that appear relatively smooth, for example geopotential at 500~hPa (figure~\ref{fig:spc_z_1}), while for other fields differences in the spectra are smaller (figure~\ref{fig:spc_t_1}).

\section{Discussion}
Optimising a  proper score objective is a powerful method to generate probabilistic forecasts of complex dynamical systems like the atmosphere \citep{lang2024aifs-crps, pacchiardi2024probabilistic, Shokar_2024, kochkov2024neural}. It enables optimising forecast skill over long forecasts, and one forecast step only requires a single model evaluation. This is in contrast to, for example, the diffusion paradigm \citep{sohldickstein2015deepunsupervisedlearningusing, ho2020denoising, song2021scorebasedgenerativemodelingstochastic}, which also has been shown to be an effective method to generate probabilist forecasts \citep{price2023gencast, ensemble_blog, larsson2025diffusionlamprobabilisticlimitedarea}. Here, the model needs to be called many times during inference for each forecast step. However, in diffusion modelling the model learns to forecast different scales due to varying level of noise added to the target state (e.g., \citet{dieleman2024spectral}). How much emphasis is put on each scale can then be adjusted by the noise schedule during training and sampling \citep{karras2022elucidating}. In proper  score optimisation, the representation of different scales is implicit. By introducing a multi-scale loss component, it becomes possible to target specific scales and, e.g. reduce spurious variability in predictions. Consistently, \citet{kochkov2024neural} find that combining  CRPS terms computed in spectral space (one for each spectral coefficient up to wavenumber 80) with their grid-point CRPS loss improves the representation of long-range correlations in forecast  with their hybrid model.

Computing the multi-scale loss at two scales incurs only marginal additional cost beyond the model's forward and backward pass. The overhead might become more significant with a large number of scales.

More work will be required to assess what the best set of hyper-parameters is for weather forecasting. For example, while variability of geopotential fields is significantly improved, there is still an offset between analysis and forecasted fields at the smallest scales. This could be an indication that more scales will improve results further. Also, different problems - for example long-range forecasting, or downscaling - might require different scale-weighting. This will be explored in future work.

\section{Conclusion}
Introducing a multi-scale loss function in proper-score-based training, such as with the almost fair continuous ranked probability score (afCRPS), improves the representation of variability in machine-learned weather forecasting models. In our experiments, forecast skill remains unchanged while the physical realism of forecast fields is enhanced. We believe that the multi-scale loss formulation will make proper-score optimization even more attractive for a wide range of prediction tasks.

\paragraph{Acknowledgments:} \textit{We acknowledge the EuroHPC Joint Undertaking for awarding this work access to the EuroHPC supercomputer MN5, hosted by BSC in Barcelona through a EuroHPC JU Special Access call. 
}

\bibliography{references}

\end{document}